\documentclass[journal]{IEEEtran}
\ifCLASSINFOpdf
\usepackage{graphicx}
\usepackage{cuted}
\else
\fi
\begin{document}
\title{A Techno-Economic Cost Framework for Satellite Networks Applied to Low Earth Orbit Constellations: Assessing Starlink, OneWeb and Kuiper}
\author{Osoro B. Ogutu
        and Edward J. Oughton% <-this % stops a space
\thanks{O. Ogutu is with the Department
of Geography and Geoinformation Science, George Mason University, Fairfax,
VA, 22030}% <-this % stops a space
\thanks{E. Oughton is an assistant professor with the Department
of Geography and Geoinformation Science, George Mason University, Fairfax,
VA, 22030 eoughton@gmu.edu}% <-this % stops a space
\thanks{}}
\markboth{A Techno-Economic Cost Framework for Satellite Networks Applied to Low Earth Orbit Constellations}%
{ \MakeLowercase{\textit{}}:}

\maketitle

\begin{abstract}
	Delivering broadband connectivity to unconnected areas is extremely challenging. The emergence of Low Earth Orbit (LEO) satellite systems has been seen as a potential solution for connecting remote areas where engineering terrestrial infrastructure is prohibitively expensive to deploy. Despite the hype around these new technologies, we still lack an open-source modeling framework for assessing the techno-economics of satellite broadband connectivity which is therefore the purpose of this paper. Firstly, a generalizable techno-economic model is presented to assess the engineering-economics of satellite constellations. Secondly, the approach is applied to assess the three main competing LEO constellations which include Starlink, OneWeb and Kuiper. This involves simulating the impact on coverage, capacity and cost as both the number of satellites and quantity of subscribers increases. Finally, a global assessment is undertaken visualizing the potential capacity and cost per user via different subscriber scenarios. The results demonstrate how limited the capacity will be once resources are spread across users in each satellite coverage area. For example, if there is 1 user per 10 km${^2}$ we estimate a mean per user capacity of 24.94 Mbps, 1.01 Mbps and 10.30 Mbps for Starlink, OneWeb and Kuiper respectively in the busiest hour of the day. But if the subscriber density increases to 1 user per km${^2}$, then the mean per user capacity drops significantly to 2.49 Mbps, 0.10 Mbps and 1.02 Mbps for Starlink, OneWeb and Kuiper respectively. LEO broadband will be an essential part of the connectivity toolkit, but the results reveal that these mega-constellations will most likely have to operate below 0.1 users per km${^2}$ to provide a service that outcompetes other broadband connectivity options. The open-source codebase which the paper contributes is provided with the hope that other engineers will access, use, and further develop the satellite assessment capability.
\end{abstract}

\begin{IEEEkeywords}
Low Earth Orbit Economic, Technoeconomic,  Broadband, Satellite
\end{IEEEkeywords}

\IEEEpeerreviewmaketitle

\section{Introduction}
\IEEEPARstart{I}{nternet} Internet connectivity is a catalyst for societal and economic development, with importance in both emerging and frontier economies \cite{b1}-\cite{b5}. Exactly how to bring unconnected communities online has been a subject of discussion for decades, leading to the rise of numerous global taskforces responsible for evaluating affordable ways of delivering universal broadband connectivity\cite{b5}\cite{b6}. Indeed, over 3 billion of the world’s population are yet to get online, while over 1 billion people are living in an area with no Internet connectivity\cite{b7}. The absence of network infrastructure is cited as major reason why so many people remain offline\cite{b8}\cite{b9}. Therefore, new engineering approaches are required to help lower deployment costs and help connect the remaining population\cite{b10}\cite{b11}.  

One of the cheapest ways to supply wide-area broadband connectivity is via cellular technologies, hence delivery in low and middle-income countries has been dominated by Mobile Network Operators (MNOs) (despite governments investing in their own High Throughput Satellite broadband capabilities)\cite{b12}\cite{b13}. However, MNOs have been experiencing challenging business conditions in recent years. Declining Average Revenue Per User (ARPU) globally has led to static or decreasing revenues, making it even harder to deploy new infrastructure in hard-to-reach areas. For instance, between 2018 and 2019 the global ARPU fell by 1\% \cite{b14}\cite{b15}. While significant progress has been made in recent years, statistics indicate that the growth rate of the number of Internet users has slowed globally, suggesting it is getting harder to add new users, often due to challenging engineering-economic conditions. This highlights the importance of engineering innovative supply-side technologies for connecting hard-to-reach users, particularly if they can overcome many of the economic barriers facing deployments in rural and remote areas.  

A variety of alternative policy and technology solutions have been proposed \cite{b16}. One of the options which has received the most media attention is LEO satellite constellations to deliver high-capacity wireless broadband connectivity and support the deployment of the Internet of Things (IoT)\cite{b17}-\cite{b19}. The aim is to increase the available data rate, achieving higher Quality of Service (QoS), thus helping to lower the cost per bit for serving the hardest-to-reach areas. Several technical developments are required to ensure these broadband services can be delivered in affordable ways, ranging from spectrum sharing to adaptive control and beamhopping \cite{b20}-\cite{b21}. This is further complemented by the launch of dense networks of cheap and mass-produced satellites that decrease the coverage area of each asset, thereby increasing the level of spectral reuse compared to other satellite systems, for example, in Geostationary Orbit (GEO). Many companies have grand ambitions for their own constellations, including SpaceX’s Starlink, OneWeb and Blue Origin’s Kuiper. Surprisingly however, there has been relatively little analysis on the potential data rates and costs involved in delivering wireless broadband connectivity via LEO constellations. For example, how does the quality of the broadband services provided by these engineered systems play out spatially across the globe? There has already been widespread interest from engineers, economists, and policy makers regarding their operation. This interest includes the challenges they may face and the potential use of these technologies in closing the digital divide, particularly how they match up with other broadband options such as 5G or IEEE 802.11ax (Wi-Fi 6)\cite{b22}-\cite{b28}. Much of the existing research focuses purely on technical engineering aspects of LEO constellations, without consideration of the per user received capacity or cost at the sub-national level in each country across the globe \cite{b29}\cite{b30}. 

Subsequently, the contribution of this paper is to examine these dimensions by developing an open-source engineering-economic simulation model. Such an approach also enables others, should they choose to do so, to access the developed codebase to reproduce the analysis and use the resulting analytics to inform their own future decisions (whether engineering, economic or policy-related). The assessment focuses on applying the approach to three LEO constellation systems, including Starlink, OneWeb and Kuiper to produce insight on (i) the potential capacity per user and (ii) the potential cost per user. The research questions are articulated as follows: 

\begin{enumerate}
        \item How much capacity can be provided by different LEO broadband constellations? 
        \item What is the potential capacity per user from different constellations?  
        \item What is the potential cost per user as subscriber penetration increases? 
        \item Which parts of the world are LEO constellations most suitable for? 
\end{enumerate}

The paper is structured as follows. Next a literature review will be carried out, followed by a description of the method in Section IV. The application of the approach to the different constellations is articulated in Section V, with the results reported in Section VI.  A discussion of the ramification of the results is undertaken in Section VII before conclusions are given in Section VIII.

\section{LEO Constellations and Broadband Connectivity}
\label{sec:headings}

Recently, there has been a shift towards LEO constellations, defined as those satellites located below the altitude of 2000 km\cite{b31}, as opposed to GEO above 35,000 km and Medium Earth Orbit (MEO) between 5,000-12,000 km\cite{b32}. Traditionally, Internet provision has been delivered through GEO and MEO satellites, but the high latency and costs has made them unpopular\cite{b33}. Mega constellations have now emerged with SpaceX promising to launch 12,000 satellites as part of Starlink, along with similar plans by OneWeb and Blue Origin, all hoping to provide globally available broadband connectivity.  

The use of LEO systems provides many engineering advantages as well as applications such as broadband provision for high speed trains and aircrafts\cite{b34}\cite{b35}. Due to the lower orbit location and novel engineering, data processing and relaying optimization techniques, data packets have shorter propagation delay quantified by a Round Trip Time (RTT) that can be as low as 100 ms\cite{b36}\cite{b37}. The relatively low RTT is tolerable for many current media applications but not for delay sensitive uses such as online video gaming, video calling or future real-time IoT\cite{b38} \cite{b39}. Secondly, LEO systems permit the usage of high frequency bands such as Ku, Ka, Q and V bands that offer large bandwidth as opposed to those for GEO satellites, meaning higher capacities can be provided to users \cite{b40}\cite{b41}.  
\par
Economically, LEO systems are more scalable than other systems (in terms of adding capacity), as a constellation can easily be added to without disrupting existing broadband services \cite{b42}. For instance, SpaceX plans to eventually add 42,000 satellites but will start with a first batch of about 5,000 before moving to 12,000. Compared to GEO, the complexity and cost per satellite in a LEO system is lower, and redundancy can continually be improved without interfering with the rest of the system. This has made them attractive for other missions such as navigation\cite{b43}\cite{b44}. On the downside, LEO systems experience high overhead costs because continuous launches are required to add more satellites, including replacing decommissioned assets (resulting from their orbit location and small life span of about five years).

Other than the cost challenges of LEO constellations, there are notable technical limitations of their usage. Most of the LEO satellites travel at speeds between 5 to 10 km/s. Consequently, users on the ground have only few minutes to connect and communicate with the satellite. This results in frequency changes (doppler shift) that can degrades QoS unless optimization algorithms and engineering modifications are made to the receiver end\cite{b45}\cite{b46}. Satellite Network Operators (SNOs) address the problem through dynamic management of radio resources to improve QoS without increasing the level of interference\cite{b47}. Additionally, the high frequencies associated with LEO satellites are extremely affected by rain attenuation especially in tropical regions where the effects are significant to as low as 7 GHz\cite{b48}\cite{b49}. This has seen SNOs designing their constellations with multiple satellites to provide redundancy because of high unavailability\cite{b50}. 
\par
However, the greatest advantage of LEO systems and satellites in general lie in their ability to serve remote areas\cite{b51}\cite{b52}. This is particularly the case in serving extreme topographies when there are cliffs, valleys, steep slopes and geologically disaster-prone areas where terrestrial networks are expensive to implement due to engineering complexities and cost implications. Delivering 5G-like services to rural and remote areas (high capacity, low latency) may not be possible via traditional infrastructure deployment due to viability issues\cite{b53}. This presents an opportunity for satellite operators in providing services to areas which are unviable with wireless or fixed broadband technologies\cite{b54}.

\section{LEO Constellations and Broadband Connectivity}
\label{sec:headings}
\begin{figure*}[!]
    \includegraphics[width=\linewidth, height=\textheight,keepaspectratio]{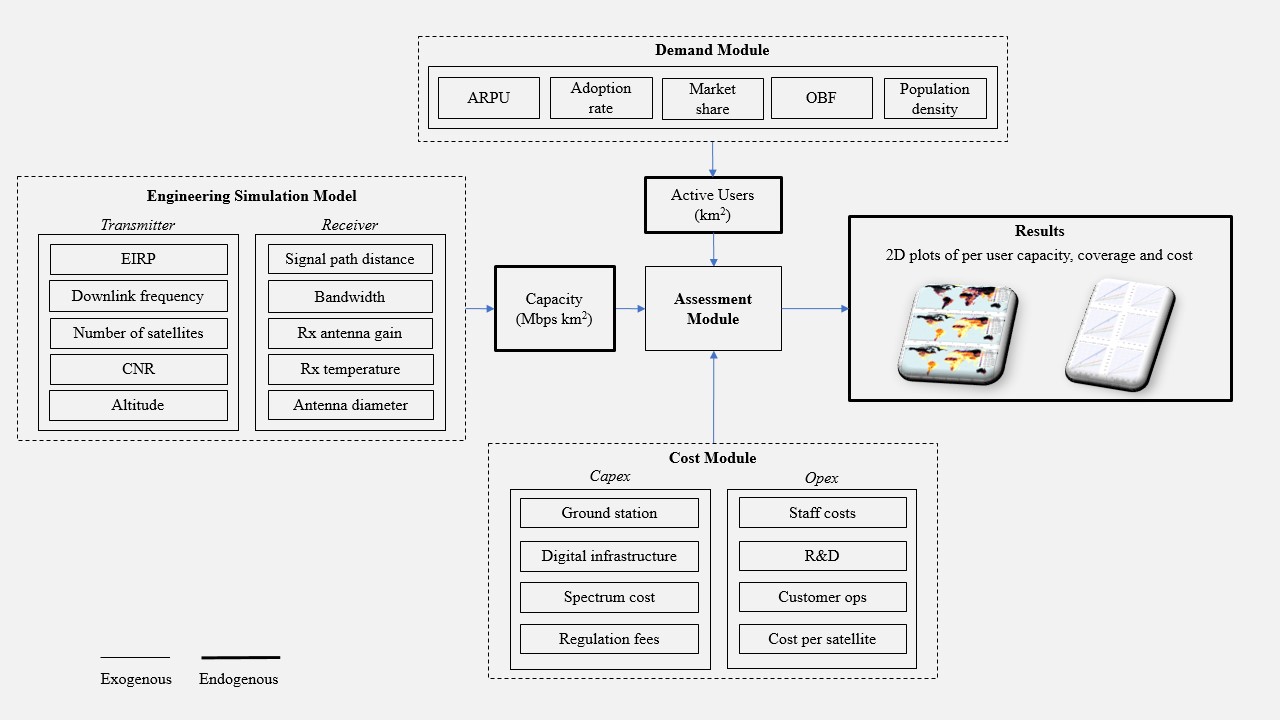}
    \caption{Engineering-economic framework for a LEO satellite network}
    \label{fig:my_label1}
\end{figure*}

In this method an engineering-economic framework is defined for a single satellite network, which in this case is focused on LEO (although the approach could easily be adapted for MEO or GEO). The open-source software repository enables users to access the model code and adapt the framework to other constellations, as desired. An overview of this framework is visualized in Figure \ref{fig:my_label1} detailing the exogenous inputs as well as the endogenously determined outputs. 

Firstly, a supply-side engineering system model is defined which captures the capacity and coverage aspects of a new constellation. Secondly, a cost model is presented which enables the total cost of operating a constellation to be estimated. 

\subsection{System model}
In the system modeling approach taken, the focus is placed on the access-side between satellites and user terminals, with emphasis on the downlink capacity. The capacity of a wireless network is dependent on three key factors which include (i) the available spectral efficiency of the radio interface (bits per Hz), (ii) the level of spectral reuse via network densification (adding more satellites to the constellation), and (iii) the quantity of available spectrum bandwidth (augmenting the total bandwidth across all channels)\cite{b76}.

The resulting QoS can also be severely affected by several physical factors, particularly geographic distance, and topography, with higher signal propagation losses translating to lower data rates\cite{b77}. We focus on modeling and simulating the system downlink capacity by estimating transmitted power, losses and the resulting Carrier-to-Noise Ratio (CNR) as a semi-random process.

\begin{figure*}[!]
    \includegraphics[width=\linewidth, height=\textheight,keepaspectratio]{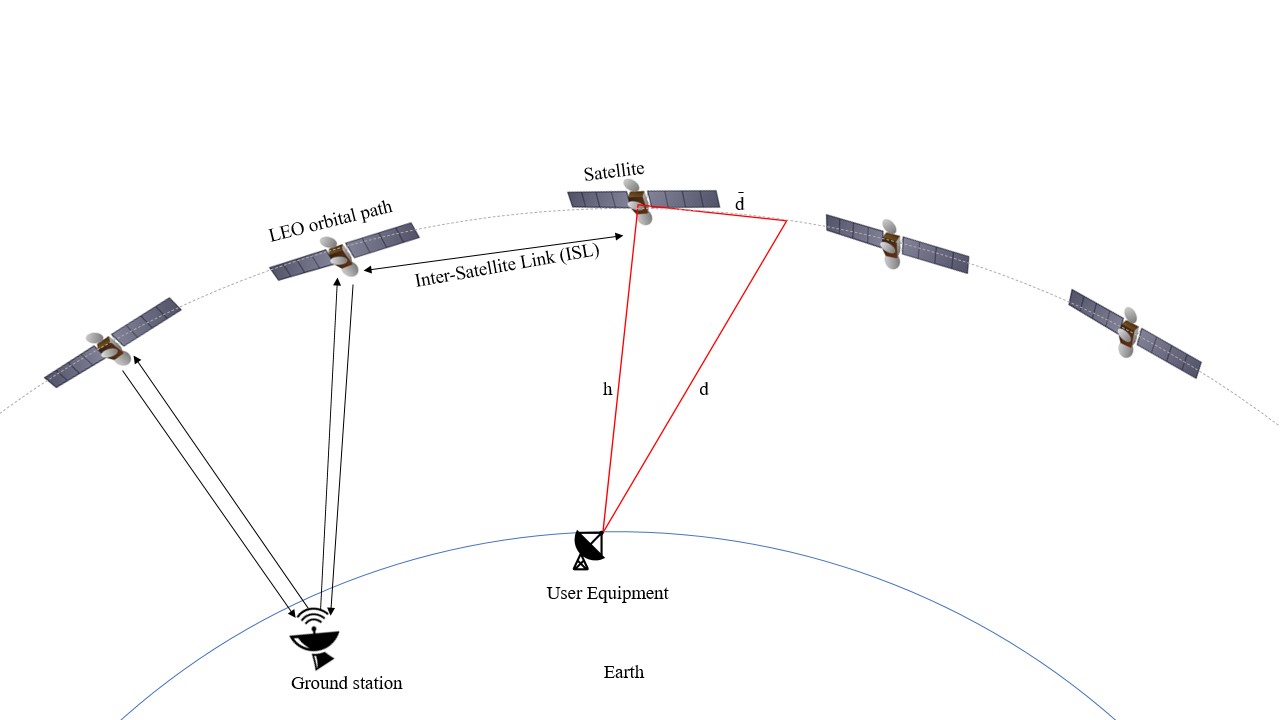}
    \caption{The geometry of the mean path length in a satellite constellation}
    \label{fig:my_label}
\end{figure*}

We make conservative estimates of capacity to avoid overestimation. Thus, the system model estimates the available capacity of the satellite network using a stochastic geometry method, based on first finding the mean path length ($d$) between the transmitter and receiver as per the hypotenuse in Figure \ref{fig:my_label}.
Given a particular constellation with a known number of satellites ($S_{n}$), the network density ($S_{Nd}$) ($km^{-2}$) can be established as follows:
\begin{equation}
    S_{Nd}=\frac{S_{n}[satellites]}{A_{Earth}[km^{2}]}
\end{equation}
Where $A_{Earth}$ is the area of the earth.
The satellite coverage area ($S_{coverage}$) ($km^{2}$) is given by:
\begin{equation}
    S_{coverage}=\frac{A_{Earth}[km^{2}]}{S_{n}[satellites]}
\end{equation}

The mean distance ($\bar{d}$) (km) between satellites in the constellation as indicated in Figure \ref{fig:my_label} can then be computed using equation (3).

\begin{equation}
    \bar{d}=\frac{\sqrt{\frac{1}{S_{Nd}}}}{2}
\end{equation}

Given the orbital altitude ($h$) ($km$) of the satellite, the stochastic signal propagation path distance ($d$) (km) as per Figure \ref{fig:my_label} is determined by Pythagoras as in equation (4).
\begin{equation}
    d=\sqrt{h^{2}+\bar{d}^{2}}
\end{equation}

\begin{figure*}[!]
    \includegraphics[width=1\linewidth,height=9cm,keepaspectratio]{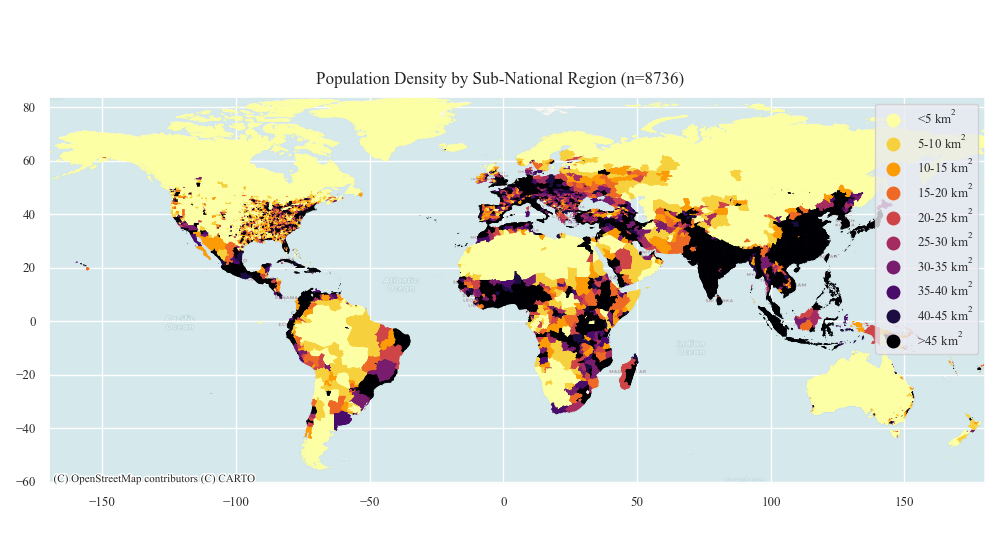}
    \caption{World population density by sub-national region}
    \label{fig:my_label2}
\end{figure*}
Similarly, the stochastic free space path loss ($FSPL$) (dB) at different satellite positions, with respect to the user terminal, is calculated by adding a pseudo-random variation using a lognormal distribution with a mean ($\mu$) of 1 and a standard deviation ($\sigma$) of 7.8 to the normal FSPL equation as shown in (5).
\begin{equation}
    FSPL=\left(\frac{4\pi df}{c}\right)^{2}+\sigma
\end{equation}
Where $f$ is the frequency of the signal being transmitted and $c$ is the speed of light, $3.0\times10^{8}$  $m/s$. We then compute the Signal-to-Noise Ratio (\textit{SNR}) for different $FSPL$. The \textit{SNR} depends on the downlink effective isotropic radiated power (EIRP) stated in equation (6). 
\begin{equation}
    EIRP=10.\log_{10}(G_{T}.P_{T})
\end{equation}
$G_{T}$ and $P_{T}$ are transmit antenna gain and power respectively. 
The figure of merit of the receiving antenna, $\frac{G_{r}}{T}$, also contributes significantly to the resultant\textit{ SNR} and is defined by equation (7). 
\begin{equation}
    \frac{G_{r}}{T}=G_{r}[dBi]+NF[dB]-10\log10(T_{0}+(T_{a}-T_{0}).10^{-0.1.NF})
\end{equation}

$G_r$ is the receiver gain, $T$ the system temperature, $NF$ is the receiver noise figure, $T_{0}$ ambient temperature and $T_{a}$ antenna temperature.
This T is correct for earth but would be lower in space, improving receiver sensitivity, potentially increasing capacity beyond our conservative estimates here.
The resultant \textit{SNR} is then obtained by equation (8).
\begin{equation}
    SNR=EIRP\left[dBW\right]+\frac{G_r}{T}\left[dBi/K\right]-FSPL\left[dB\right]
\end{equation}
$$-{OT}_{Loss}\left[dB\right]-10.\log_{10}{\left(k.T.B\right)[K]}$$
Where ${OT}_{Loss}$ is the sum of all other losses, $k$ the Boltzmann constant ($1.38064852\times10^{-23}$ $m^{2} kg s^{-2} K^{-1}$) and \textit{B} the available spectrum bandwidth. Then using the spectral efficiency ($SE$) values for different achieved SNR rates, based on the European Telecommunications Standards Institute (2020-08) documentation\cite{b78}, the resulting channel ($C_{Mbps}$) and area capacities ($C_{A}$) are calculated as per equation (9) and (10) in $Mbps$ and $Mbps/km^{2}$, respectively (\textit{ch} is the number of channels and \textit{k} the frequency reuse factor).
\begin{equation}
    C_{Mbps}=SE\times B\times ch\times k
\end{equation}
\begin{equation}
    C_{A}=\frac{C_{Mbps} [Mbps]}{S_{coverage} [km^{2}]}
\end{equation}
Using equations (9) and (10), the potential user capacity for different spatial statistical units can be approximated. This approach forms the basis for evaluating the possible capacity per user for the constellation given the number of satellite assets in orbit. 

\subsubsection{Cost model}
An overview is now provided of a cost model for a satellite constellation providing broadband connectivity. The costs of launching and operating a satellite network consist of capital expenditure (capex) for upfront investments, and then ongoing annual costs classified as operational expenditure (opex). Our aim is to obtain the discounted Net Present Value (NPV) over the chosen study period, to represent the Total Cost of Ownership (TCO) for each constellation, and thus the cost per user. The capex required to launch a constellation can be defined as follows in equation (11).
\begin{equation}
    Capex=C_{Launch}+C_{Station}+C_{Spectrum}+C_{Integration}
\end{equation}

Where $Capex$ is equivalent to the sum of satellite launching costs $(C_{Launch})$, the sum of all ground station building costs $(C_{Station})$, the spectrum acquisition costs $(C_{\ Spectrum})$ and finally any costs for integration of the system into existing terrestrial infrastructure $(C_{Integration})$. 

Moreover, the annual opex for the constellation can be defined as in equation (12):
\begin{equation}
    Opex=O_{GS\ Energy}+O_{Acquisition}+O_{RD}+O_{Lab}+ O_{Maint}
\end{equation}

Where $Opex$ is the sum of energy costs for the ground stations $(O_{GS\ Energy})$, acquisition of subscribers $(O_{Acquisition})$, research and development $(O_{RD})$, labour costs $(O_{Lab})$ and the maintenance costs $O_{Maint}$. 
From the cost parameters, we calculate the Total Cost of Ownership (TCO) estimated for each satellite asset (AssetNPV) by computing the NPV over a 5-year period (Y) at 5\% discount rate (r) illustrated in equation (13).

\begin{equation}
    ASSET_{NPV}=Capex+\sum_{t=0}^Y \frac{Opex}{(1+r)^t}
\end{equation}

Once the NPV for each asset is obtained, it is then possible to begin to connect the engineering and economic models presented so far, into an integrated techno-economic framework. 

\subsection{Integration of System and Economic Model}
After the specification of the system and cost model has been setup the two can be linked, to establish the cost of delivering broadband for different scenarios. Since the satellite’s medium access control (MAC) layer does not split capacity linearly among users in a coverage area, we set the model for different user adoption rates with an overbooking factor of 20\cite{b79}. The overbooking factor implies that 1 in 20 users access the network at the peak hour. The adoption rates are based on three take-up scenario of 0.5\%, 1\% and 2\% which is a common way to assess infrastructure demand\cite{b79}\cite{b80}. The users per square kilometer $(User_{sq\_km^{2}}$) is then established as follows. 
\begin{equation}
    User_{sq\_km^{2}}=Pop._{density}\left(\frac{Adop._{rate}}{100}\right)
\end{equation}
Where $Pop._{density}$ is the population density at sub-national regions obtained from WorldPop 2020 raster layer\cite{b81} and $Adop._{rate}$; is the adoption rate that can be equated to any of three scenarios (0.5\%, 1\% and 2\%).The estimated active users $User_{active}$ are then determined by equation (15).
\begin{equation}
    User_{active}=\frac{User_{sq\_km^{2}}}{OBF}
\end{equation}
Where $OBF$ is the overbooking factor set at 20 for the model.Using equation (10) the per user capacity ($ C_{per\_user}$) is calculated as follows:
\begin{equation}
    C_{per\_user}=\frac{CA}{User_{active}}
\end{equation}
The cost per user ($Cost_{per\_user}$) can therefore be computed using equation (17).
\begin{equation}
    Cost_{per\_user}=\frac{ASSET_{NPV}}{S_{coverage}\times User_{active}}
\end{equation}
Where $S_{coverage}$ and $SSET_{NPV}$ are obtained from the system and cost model equations (2) and (13) respectively. 

\section{Applications}
In this section we describe how we apply this framework to three LEO constellations, including Starlink, OneWeb and Kuiper over a study period of 2020-2025. We assume the constellations are already at or beyond critical coverage point and that there is enough capacity between the ground station and the satellite. We then obtain engineering parameters for the simulation with 100 iterations from public International Telecommunication Union (ITU) fillings as reported in Table 1\cite{b82}\cite{b83}. The mass of Kuiper satellite is assumed since they are yet to be launched. The final channel capacity ($C_{Mbps}$) is obtained by multiplying the total bandwidth with the frequency reuse factor obtained from previous research\cite{b67}. Since the existing systems launched such as Starlink’s v1.0 satellites are using the bent-pipe architecture, a similar bandwidth on the feeder and user links is assumed. This constrains the frequency reuse factor to the product of the polarization and the number of active feeder links resulting into a value of 2 for Starlink. However, this is likely to change with SpaceX’s announcement of more capable v2.0 satellites. A similar frequency reuse factor is assumed for both OneWeb and Kuiper satellites.  
\begin{table}[!]
\caption{Engineering parameters by LEO constellation}
\label{table}
\setlength{\tabcolsep}{3pt}
\begin{tabular}{|p{90pt}|p{35pt}|p{25pt}|p{35pt}|p{25pt}|}
\hline
\textbf{Parameter}& 
\textbf{Starlink}& 
\textbf{Kuiper}&
\textbf{OneWeb}&
\textbf{Unit}
\\
\hline
Planned Satellites& 
4,425& 
3,236&
720&
- \\
Satellite Mass&
260&
260&
147.5&
kg \\
Simulated Satellites& 
5,000& 
3,236&
720&
- \\
Downlink Frequency& 
13.5& 
17.7&
13.5&
GHz \\
Bandwidth& 
0.25& 
0.25&
0.25&
GHz \\
Channels& 
8& 
8&
8&
- \\
Aggregate Bandwidth& 
2& 
2&
2&
GHz \\
System Temperature& 
290&
290& 
290& 
K\\
EIRP& 
67.7& 
73.1& 
68.3&
dBm \\
Receiver Antenna Gain& 
37.7& 
43.1& 
38.3& 
dBi\\
Altitude& 
550& 
1,200&
610&
km\\
Minimum Elevation Angle& 
40& 
55& 
35.2& 
Deg\\
Antenna Diameter& 
0.7& 
1& 
0.75& 
m\\
Modulation Scheme& 
16& 
16& 
16&
APSK
\\
Frequency Reuse Factor&
2&
2&
2&
-\\
\hline
\end{tabular}
\label{tab1}
\end{table}
\begin{table}[!]
\caption{Cost parameters by LEO constellation \textit{(All values in million US dollars)}}
\label{table}
\setlength{\tabcolsep}{3pt}
\begin{tabular}{|p{90pt}|p{35pt}|p{25pt}|p{35pt}|p{25pt}|}
\hline
\textbf{Parameter}& 
\textbf{SpaceX (US\$ Millions)}& 
\textbf{Kuiper (US\$ Millions)}&
\textbf{OneWeb (US\$ Millions)}&
\textbf{Type}
\\
\hline
Ground Station& 
81.2& 
33&
47&
Capex\\
Digital Infrastructure& 
6.2& 
3.6&
2.5&
Capex\\
Spectrum Cost& 
125& 
125&
125&
Capex\\
Regulation Fees& 
0.7&
0.7& 
0.7& 
Capex\\
Cost of Operational Staff& 
60& 
60& 
7.5&
Opex\\
Cost Overhead per R\&D& 
7.5& 
7.5& 
7.5& 
Opex\\
Marketing and Customer Acquisition Cost& 
50& 
50&
50&
Opex\\
Launch Cost Per Satellite& 
0.5& 
1.5& 
2.0& 
Opex\\
Cost of each Satellite& 
0.25& 
0.25& 
0.25& 
Opex\\
Lifespan of each Satellite& 
10& 
10& 
10& 
Opex\\
\hline
\end{tabular}
\label{tab1}
\end{table}

As the LEO satellite capex and opex are not explicitly known due to commercial sensitivities, estimated values are sourced from the literature and inferred from established GEO satellite companies with publicly available financial statements\cite{b84}. Starlink has the highest aggregate cost due to the number of satellites in orbit compared to Kuiper and OneWeb. However, the costs are expected to reduce since Starlink has promised to have Inter-Satellite Link (ISL) for v0.9 satellites, just like OneWeb and Kuiper, which will reduce the need for many gateway stations \cite{b85}.

Notably, there is major difference in launch costs for the three constellations. The US\$0.5 million launch cost per satellite is based on the US\$28 million total for every launch as stated by Starlink\cite{b86}. The launch cost for Starlink is expected to be lower than the other systems since its major launching vehicle, Falcon 9 has already made 122 successful launches by 2021 with fewer than 5 failures. This drives down the total launch costs due to decreased insurance premiums, compared to Kuiper’s launch vehicles that are yet to send its first batch of satellites to LEO. The launch cost value for Kuiper is assumed from NASA’s 2018 Ames Research Centre publication which sets the launch cost to LEO at US\$90 million. Assuming 60 satellites per launch, a US\$1.5 million value per satellite is reached. For OneWeb, the US\$2 million launch cost per satellite is obtained by multiplying the satellite mass (147kg) by the US\$13,100 launch cost/kg for LEO missions as indicated in\cite{b87}. The costing uses 36 satellites per mission as already witnessed in the previous OneWeb launches. A summary of the capex and opex are shown in Table 2.

Plots of key engineering and economic metrics are produced to answer the research questions articulated earlier in this paper. The results are then broken down globally to provide insight into the capacity at the sub-regional level. We obtain layer 0, 1 and 2 boundaries for all countries from the Global Administrative Areas database to help visualize areas where LEO broadband could be most suitable\cite{b88}. We exclude countries with small boundaries such as Luxembourg for simplification. The population density (pop/km$^{2}$), area and population for all sub-national regions globally is extracted from the WorldPop 2020 raster layer\cite{b81}. 

\section{Results}
This section details the engineering and economic results. For consistency across the three constellations, we present the simulation outputs for the first 1,000 satellites for Starlink and Kuiper, comparative to the planned 720 satellites for OneWeb. The FSPL contributes the highest loss and the subsequent signal received by the users on the ground. Starlink records the lowest mean FSPL of 172.4±0.05 dB, relative to Kuiper (176.0±0.08 dB) and OneWeb (179.3±0.13 dB). The smaller FSPL recorded by Starlink is due to the lower orbital altitude of 550 km and a high minimum elevation user angle of 40°, minimizing link budget losses. Kuiper compensates its high orbital altitude (1,200 km) by also having a large minimum user elevation angle of 55°. In contrast, OneWeb's low minimum elevation of 35.2° and higher orbital altitude (610 km) results in larger path losses. This is further affected by the low density of satellites in the network leading to longer path distances, \textit{d} as defined in the system model. 
\begin{figure*}[!]
  \centering
  \includegraphics[width=\linewidth, height=\textheight,keepaspectratio]{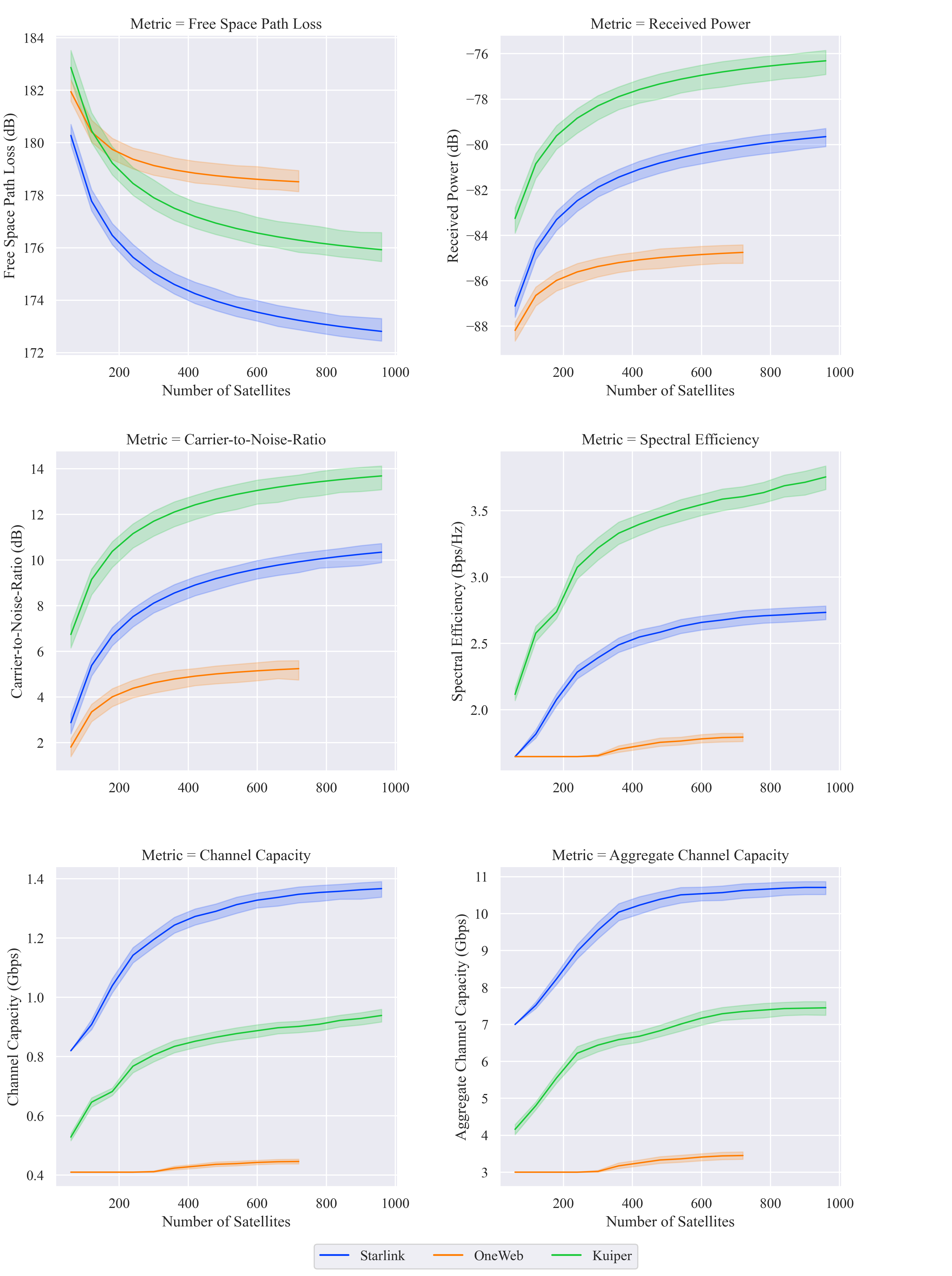}
  \caption{Engineering results for the three constellations}
  \label{fig:my_label3}
\end{figure*}
\par
The FSPL has a consequential impact on the received power for LEO satellites. However, Kuiper records the highest received power followed by Starlink and then OneWeb. The antenna design differences in the constellations result in this variation. For example, Kuiper has the highest receiver antenna gain (38.3 dBi) and diameter (0.75 m), the parameters are not sufficient to offset the larger FSPL, leading to the lowest power received. Assuming that the other noise and interference sources remain uniform across the systems, the CNR is directly proportional to the power received. This results in the mean CNR of 10.74±0.05 dB, 4.47±0.13 dB and 13.64±0.08 dB for Starlink, OneWeb and Kuiper systems, respectively. Generally, Starlink provides the best performance. 

\begin{figure*}[!]
  \centering
  \includegraphics[width=\linewidth, height=\textheight,keepaspectratio
  ]{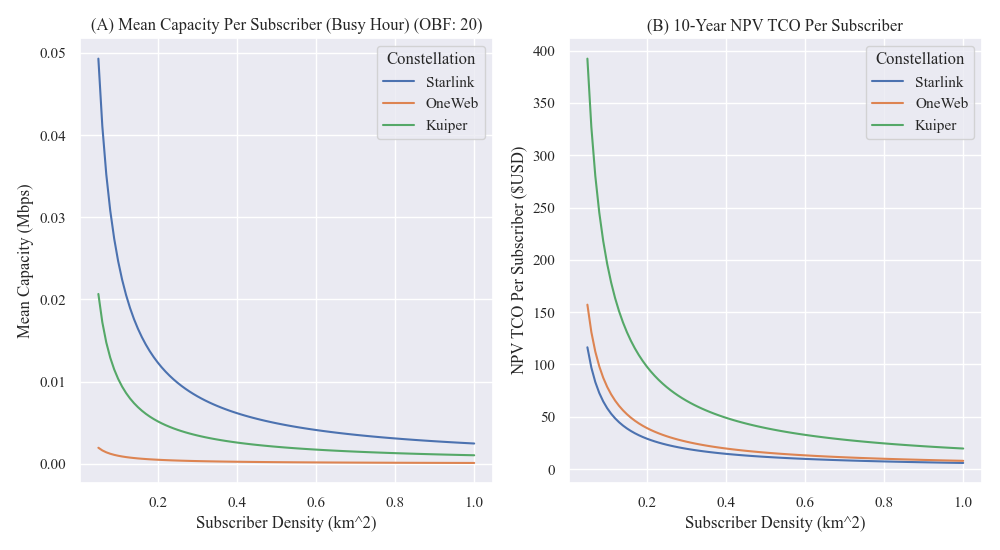}
  \caption{Mean capacity and cost metrics by subscriber density}
  \label{fig:my_label4}
\end{figure*}
\par
In Figure \ref{fig:my_label4}, we present the results of modeling the mean busy hour capacity based on a remote rural area with a subscriber density between 0.05-1 subscribers per km${^2}$. We use the highest network density of approximately 5,000, 3,200 and 720 satellites for Starlink, Kuiper and OneWeb respectively. While each constellation can provide impressive aggregate channel capacity, the available capacity needs to be shared across users in very large coverage areas (see Figure \ref{fig:my_label2} for realistic global insight on population density). So, if there are 0.1 users per km${^2}$ (so 1 user every 10 km${^2}$), the mean per user capacity of 24.94±0.72 Mbps, 1.01±0.02 Mbps and 10.30±0.25 Mbps are recorded for Starlink, OneWeb and Kuiper, respectively. This contrasts with 5 users per km${^2}$, where the provided service would essentially be unavailable for OneWeb while Kuiper and Starlink each register 0.21±0.01 Mbps and 0.50±0.01 Mbps, respectively. 
\par
On computation of the aggregate system capacity, 11.72±0.04 Gbps, 3.43±0.01 Gbps and 7.53±0.03 Gbps are recorded for Starlink, OneWeb and Kuiper constellations respectively. The first two values compare to previous estimates \cite{b67} of 10 Gbps and 5 Gbps for Starlink and OneWeb respectively. The 5 Gbps recorded in the literature for OneWeb takes into account the issue of ISL. A comparison for Kuiper is not possible since previous studies have not simulated this constellation \cite{b67}. Futhermore, there are differences in the capacities recorded by Starlink in the literature with those estimated here due to the number of satellites presented in Figure \ref{fig:my_label3}. The results of only 1,000 satellites are presented in the figures. 
\par 
At maximum network density, each Starlink satellite covers approximately 100,000 km${^2}$, OneWeb 708,000 km${^2}$ and Kuiper 158,000 km${^2}$. At a subscriber density of 0.05 users per km${^2}$, the corresponding number of subscribers per satellite for Starlink, OneWeb and Kuiper are 5,000, 35,400 and 7,900 respectively. Since the aggregate capacity is shared among the subscribers, Starlink provides the highest mean capacity followed by Kuiper and OneWeb as shown in Figure \ref{fig:my_label3}. Therefore, an increase in population density (and logically a higher subscriber density) leads to a drastic decrease in mean capacity.  
\par
We also plot the potential cost in Figure \ref{fig:my_label4}. The NPV for a single satellite asset over the study period was estimated at US\$0.6 million, US\$5.6 million, and US\$3 million for Starlink, OneWeb and Kuiper, respectively. Thus, the NPV cost per user for each constellation can then be plotted which logically reduces as each subscriber density increases. Starlink incurs the least cost per user over the study period (2020-2025) that ranges US\$100-US\$10 for the subscriber density range of 0.005-1.0 (km${^2}$). Kuiper records the largest cost per user ranging between US\$400 and US\$30 for the same subscriber density range. The important caveat to these estimates is that there would be a major impact on the capacity available for each subscriber at the maximum adoption rate, due to increased contention. Hence, active constellations such as Starlink have already begun limiting adoption in high demand areas, to ensure QoS can be guaranteed to existing customers, ensuring the available broadband services remain competitive against competing technologies. 
\par 
Figure 3 illustrates population density globally by sub-national region for population deciles ranging from below 5 people per km${^2}$, to over 45 people per km${^2}$. These decile boundaries were selected because we know a priori that higher density areas will be less suitable for LEO broadband constellations, and that they will be focusing on the bottom 5\% of the market not currently served by conventional terrestrial broadband services using either fixed or wireless technologies.
\par 
We can see large parts of Asia (India, China etc.) will be unsuitable, along with most of mainland Europe (e.g. Germany, Italy) and central America (e.g. Mexico). However, the constellations can choose to limit the number of subscribers in such regions to provide relatively higher speeds and ensure QoS. In the USA, the West and South West have large areas which could be suitable, along with much of Canada, Australia and New Zealand. 
\par 
In South America large parts of the Amazon may also have low enough population density to be suitable, as well as much of the Sahara region in Africa, although whether incomes would enable the purchasing of such services would be a main concern.  
\par 
Therefore, to explore the suitability of these constellations we use a 1\% adoption rate among the local population to explore capacity per user in the busiest hour of the day. Starlink provides impressive capacity for remote regions with global coverage thanks to its high asset density. In regions with very low population density Starlink provides a mean of over 90 Mbps per user, such as in parts of Canada, the West and South West of the USA, Central and South America, Sahara Africa, South-west Africa, Australia, Russia and remote parts of Asia. Kuiper performs similarly, with only slightly reduced performance. However, OneWeb offers generally lower capacity per user, although still reaching impressive peak rates in areas with very low population density. 
\begin{figure*}[!]
  \centering
  \includegraphics[width=\linewidth, height=\textheight,keepaspectratio]{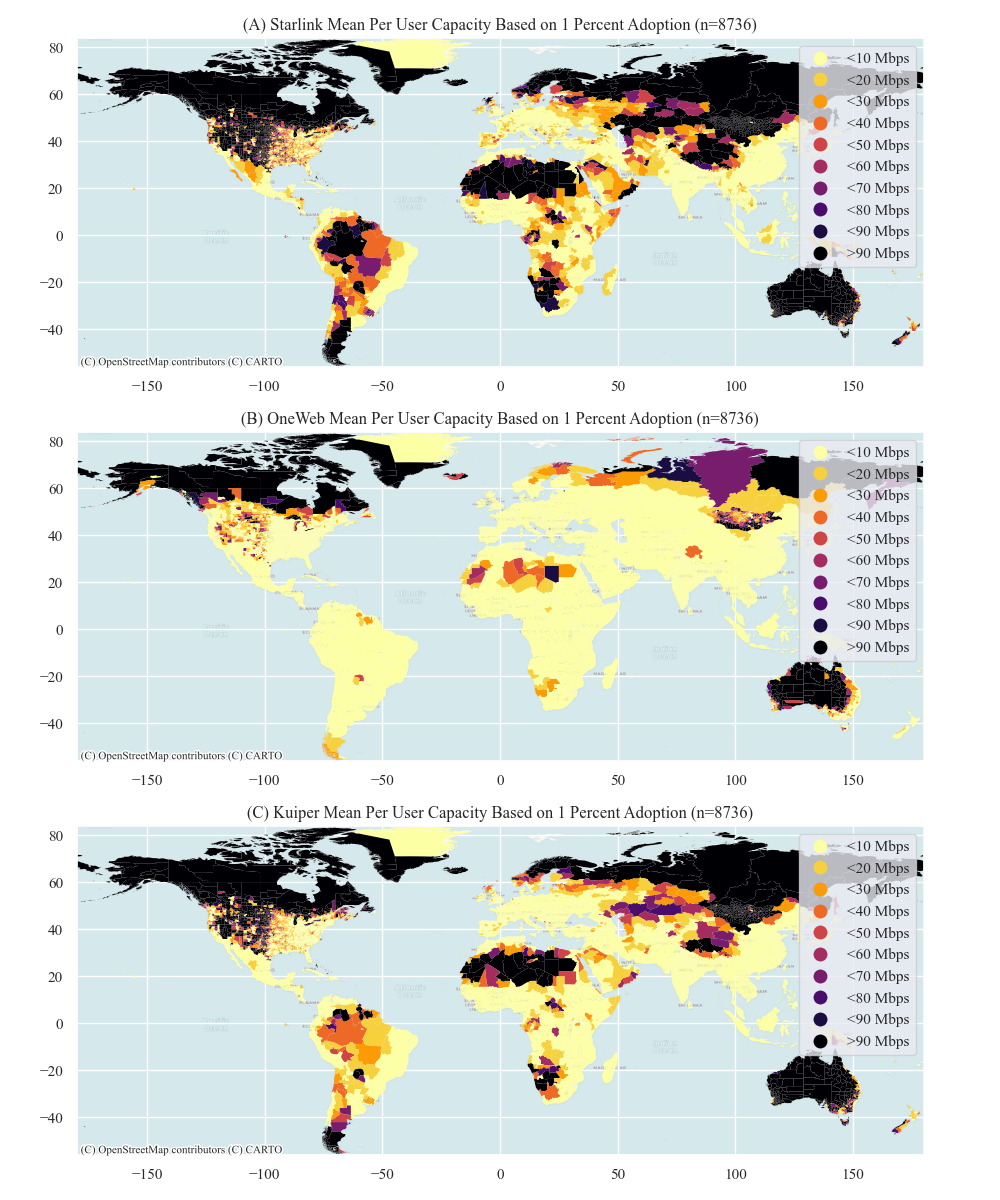}
  \caption{The per-user capacity for the three constellations in different sub-national regions of the world}
  \label{fig:my_label7}
\end{figure*}

\section{Discussion}
In this paper a generalizable techno-economic assessment model was developed for satellite broadband constellations. The approach was used to estimate the capacity and related costs for three LEO constellations, including Starlink, OneWeb and Kuiper. The open-source codebase is provided to help boost scientific reproducibility, as well as support other engineers or business analysts working in this research area\cite{b87}. The method consisted of a mix of engineering simulation, cost estimation and Geographical Information System (GIS) techniques, combined to provide new insight into the per user capacity and cost. Such analytics are very useful to help narrow the broadband availability gap in rural and remote areas by providing geospatial insight on the suitability of these technologies. The results demonstrate the connectivity opportunities and constraints of different LEO systems, as well as their viability. This section now revisits the research questions posed in the introduction of the paper. The first research question was articulated as follows: 

\subsection{How much capacity can be provided by different LEO broadband constellations?}
The findings support existing theory whereby the capacity provided by the constellation is a function of the number of satellites. Fewer satellites result in a larger coverage area and vice versa. Unlike GEO, a satellite located at LEO will also have a shorter path length. As more satellites are added into the constellation, the coverage area per satellite reduces. Furthermore, the instantaneous number of satellites available to a ground user increases. We find that for network densities of 5,000, 720 and 3,236 satellites for  Starlink, OneWeb and Kuiper respectively, the estimated coverage areas equate to 101,000, 708,000 and 157,000 km${^2}$.  
\par 
The variation in the FSPL due to the orbital altitude and network density among the three constellations results in different received power. To compensate for high path loss, Kuiper and OneWeb opt for high receiver antenna gain, transmitted power and diameter. In contrast, the ultra-dense network and low orbital altitude enables Starlink to maintain large minimum elevation angles for its users compared to the other three systems, leading to superior QoS. This explains the constellation’s Business-to-Consumer (B2C) approach as users can easily connect to its satellites with minimum engineering requirements. In contrast, the limited capacity demonstrated in this analysis for OneWeb suggests why a more enterprise-focused approach is being adopted to provide Business-to-Business (B2B) global connectivity services, ranging from cellular backhaul to logistics to emergency services redundancy.
\subsection{What is the potential capacity per user from different constellations?}
Related to the previous question, the per user capacity is therefore also positively correlated with the increase in the number of satellites for each constellation. The highest mean user capacity is achieved with the lowest subscriber densities, which occur in the most rural and remote regions where network contention is at its lowest. For instance, with 1 user every 10 km${^2}$ (0.1 users per km${^2}$) the best performing constellation (Starlink) records a very modest mean per user capacity of 24.94±0.72 Mbps. This is worse for Kuiper and OneWeb with 10.30±0.25 Mbps and 1.01±0.02 Mbps respectively. Hence, this explains why LEO broadband providers have been making a strong business case for the usage of satellites in the final 3 percent of customers in the hardest-to-reach rural and remote regions of the USA, Canada, United Kingdom, Australia and New Zealand (among other countries) due to their competitive advantage in these challenging deployment situations. While the aggregate speeds estimated are impressive, each satellite asset can easily become saturated, especially in higher populated urban and suburban areas, meaning SNOs will have to strictly manage spatial adoption rates. There is no doubt that the potential speeds per user which could be provided are highly desirable (and indeed revolutionary) for users which have struggled to gain a decent broadband connection from traditional providers. The potential services available would be more than adequate to enable intensive applications such as High Definition video streaming without buffering (providing QoS was well managed).
\subsection{What is the potential cost per user as subscriber penetration increases?}
The largest capital expenditure costs are incurred by rocket launches, building ground stations and acquiring spectrum. As more satellites are launched, the cost per user would increase, partly due to the rising operating costs, but this would ensure a better QoS for each user terminal thanks to smaller coverage areas with fewer shared spectrum resources. With more satellites in each constellation, the ground station energy requirements, maintenance, continual engineering, staff costs and general research and development increases. At a low subscriber density, high capacity per user is available but the cost could be prohibitively expensive for some. In contrast, at a high subscriber density, the cost of broadband connectivity services is much more affordable but there is a major trade-off in QoS, with only very modest speeds being delivered. 
\par 
The results open a question on whether LEO constellations could break into the urban broadband market given that MNOs and other operators can offer the services at a lower cost per user. While acquiring a segment of the urban market cannot be ruled out, the possibility of succeeding in developed countries where constellations such as Starlink are testing their products is low (driven by the need to limit the number of active users). Consequently, LEO broadband systems are more likely to play a significant role in providing global communications for niche industrial activities which require substantial mobility with high reliability. For example, maritime, rail, aviation and integration into other supply chain IoT architectures, thanks to LEO pole-to-pole coverage. Furthermore, LEO systems might also have a useful niche in delay sensitive applications such as monitoring offshore solar and wind farms in smart grid applications, thanks to the lower latency they can achieve relative to other technologies such as GEO. Alternatively, LEO broadband constellations can present a viable cost-effective solution for developing countries with growing urban centers that are yet to enjoy decent cellular and fiber infrastructure availability. However, this very much depends on the necessary spectrum being allocated in appropriate bands by each telecommunications regulator. 
\subsection{Which parts of the world are LEO constellations most suitable for?}
The performance of the three constellations in areas of different population density shows a general trend. Regions with low population density generally experience higher capacity per user with Starlink and Kuiper providing superior speeds. 
\par 
The simulation of possible geographical areas of adoption indicates that most parts of Central Asia, Middle East, South East Asia, South America, Sub-Saharan Africa and Eastern Europe are less suitable for LEO constellations with quite low capacity provided (below 10 Mbps) using the modeling parameters explored.  
\par 
These results are arrived at by only considering population density. Future research should recognize the roles of adoption factors such as disposable income, perceived relevance of the Internet, literacy and cellular network penetration, as these may affect the number of people who can actually afford to pay for broadband services.

\section{Conclusion}
Connecting the global population who are still unable to access a decent broadband service remains a key part of the United Nation's Sustainable Development Goals (specifically Target 9.c).  
\par 
Motivated by these developments, the framework applied in this paper introduces a techno-economic modeling approach for the integrated assessment of data capacity and investment cost per user by constellation. The model presents the engineering and economic simulation results using a single framework, unlike other approaches where this may be undertaken by two separate groups of professionals (engineers and business analysts).This theoretical model allows for estimation of the constellation capacity based on the known engineering parameters filed with local or global regulatory authorities such as Federal Communication Commission (FCC) and ITU. Using the information publicly available from such organizations, and estimation based on financial statements filed by publicly traded GEO, MEO and LEO broadband companies, the values can be imputed in the model to approximate the capacity and cost of delivering satellite Internet. The model has been tested for three different constellations with varying number of simulated satellites (720, 3236 and 5000) to derive the per user capacity and costs. The codebase for the model is fully open-source and available from the online repository, enabling anyone to access and further enhance the capability developed \cite{b89}.
Future research could include addressing the issue of non-linearity in the multiple access of satellite resources, which would improve on existing simplifications. Moreover as the modelling approach is generalizable for satellite constellations, the framework can be further adapted for other planned constellations, such as Telesat.
\par 
The results of the model reveal that at the 95\% confidence level, mean aggregate capacity speeds of 11.72±0.04 Gbps, 3.43±0.01 Gbps and 7.53±0.03 Gbps are achievable for Starlink, OneWeb and Kuiper, respectively. The current anticipation associated with the benefits of LEO broadband constellations is very high, but success will depend on maintaining relatively low spatial subscriber densities, preferably below 0.1 users per km${^2}$ (so less then 1 user per 10 km${^2}$), otherwise the services provided may offer little benefit against other terrestrial options. For example, the model has shown that at 0.1 users per km${^2}$, only a mean per user capacity of 24.94±0.72 Mbps, 1.01±0.02 Mbps and 10.30±0.25 Mbps can be achieved by Starlink, OneWeb and Kuiper respectively in the busiest hour of the day.
\par 
Future research needs to combine the use of this estimation method for LEO constellations, include a more sophisticated link layer model with other global cellular and fiber models, to estimate the most suitable technology for each sub-national region, based on the available demand and cost of supply.

\section*{Acknowledgment}

The authors would like to thank George Mason University for supporting the research, as well as Julius Kusuma and Jonathan Brewer for informal peer-review feedback.

\ifCLASSOPTIONcaptionsoff
  \newpage
\fi


\begin{IEEEbiography}[{\includegraphics[width=1in,height=1.55in,clip,keepaspectratio]{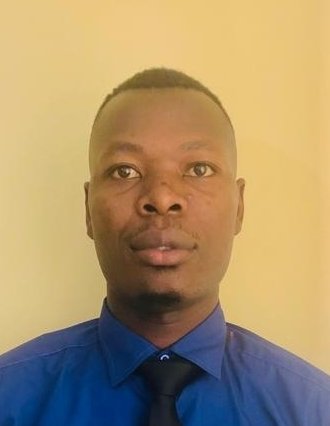}}]{\textbf{Osoro B. Ogutu}} Received his B.ENG degree in Geospatial Engineering from The Technical University of Kenya in 2017 and MSc. Degree in Satellite Applications with Data Science from University of Strathclyde, Glasgow, United Kingdom in 2020. From 2017 to 2018 he was trained by the Development in Africa with Radio Astronomy (DARA) on basics of radio antenna operation, application, signal processing, analysis, and reduction of data. His research interests include signal transmission to LEO, particularly for 5G, and the application of data science and artificial intelligence techniques in satellite data.
\end{IEEEbiography}

\begin{IEEEbiography}[{\includegraphics[width=0.8in,height=1.25in,clip,keepaspectratio]{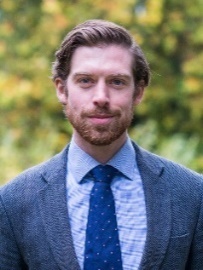}}]{\textbf{Edward J. Oughton}} Received the M.Phil. and Ph.D. degrees from the Clare College, University of Cambridge, U.K., in 2010 and 2015, respectively. He later held research positions at both Cambridge and Oxford. He is currently an Assistant Professor with George Mason University, Fairfax, VA, USA, developing open-source research software to analyze digital infrastructure deployment strategies. He received the Pacific Telecommunication Council Young Scholars Award in 2019, Best Paper Award 2019 from the Society of Risk Analysis, and the TPRC Charles Benton Early Career Award 2021. 
\end{IEEEbiography}


\begin{thebibliography}{1}

\bibitem{b1} S. Ghosh, "Broadband penetration and economic growth: Do policies matter?", \emph{Telematics and Informatics}, vol. 34, no. 5, pp. 676-693, 2017. Available: 10.1016/j.tele.2016.12.007. 

\bibitem{b2}M. Araújo, L. Ekenberg and J. Confraria, "Satellite backhaul for macro-cells, as an alternative to optical fibre, to close the digital divide," \emph{2019 IEEE Wireless Communications and Networking Conference (WCNC)}, Marrakesh, Morocco, 2019, pp. 1-6, doi: 10.1109/WCNC.2019.8885759.

\bibitem{b3} L. Townsend, C. Wallace, G. Fairhurst and A. Anderson, "Broadband and the creative industries in rural Scotland", \emph{Journal of Rural Studies}, vol. 54, pp. 451-458, 2017. Available: 10.1016/j.jrurstud.2016.09.001.

\bibitem{b4} M. Khaturia, P. Jha and A. Karandikar, "Connecting the Unconnected: Toward Frugal 5G Network Architecture and Standardization", \emph{IEEE Communications Standards Magazine}, vol. 4, no. 2, pp. 64-71, 2020. Available: 10.1109/mcomstd.001.1900006.

\bibitem{b5}Broadband Commission for Sustainable Development, ITU, UNESCO. (2018). Broadband targets for 2025. Technical report international telecommunication union accessed:  Accessed: 15- Feb- 2021, [Online] Available:
\underline{https://www.gsmaintelligence.com/research/}  

\bibitem{b6} K. Bahia and S. Suardi, The State of Mobile Internet Connectivity 2019. \emph{GSMA Intelligence}, 2019.

\bibitem{b7} J. Garrity and A. Garba, The Last-mile Internet Connectivity Solutions Guide Sustainable connectivity options for unconnected sites. \emph{International Telecommunication Union Development Sector}, 2020.

\bibitem{b8} L. Townsend, S. Arjuna, F. Gorry and W. Claire, “Enhanced broadband access as a solution to the social and economic problems of the rural digital divide.,” \emph{Local Economy}, vol. 28, no. 6, pp. 580-595, 2013.

\bibitem{b9} P. Bondalapati et al., "SuperCell: A Wide-Area Coverage Solution Using High-Gain, High-Order Sectorized Antennas on Tall Towers", \emph{arXiv preprint arXiv:2012.00161.}, 2020.

\bibitem{b10}C. L. Stergiou, K. E. Psannis, and B. B. Gupta, ‘IoTbased Big Data secure management in the Fog over a VOLUME XX, 2021 18 6G Wireless Network’, \emph{IEEE Internet of Things Journal}, pp. 1–1, 2020, doi: 10.1109/JIOT.2020.3033131.

\bibitem{b11}C. Huang et al., ‘Holographic MIMO Surfaces for 6G Wireless Networks: Opportunities, Challenges, and Trends’, \emph{IEEE Wireless Communications}, vol. 27, no. 5, pp. 118–125, Oct. 2020, doi: 10.1109/MWC.001.1900534.

\bibitem{b12} E. Oughton, "Policy options for digital infrastructure strategies: A simulation model for broadband universal service in Africa", \emph{	arXiv:2102.0356}, 2021

\bibitem{b13} Y. Panggau and M. Asvial, "Analysis of Satellite Broadband Access Implementation in Indonesia Using Costing Methodology", \emph{The 2018 International Conference on Control, Electronics, Renewable Energy and Communications (ICCEREC)}, pp. 30-35, 2021.

\bibitem{b14} \emph{Data - GSMA Intelligence.}:GSMA Intelligence, Accessed on: Accessed: 15- Feb- 2021, [Online] Available:
\underline{https://www.gsmaintelligence.com/data/}

\bibitem{b15} "Broadband Maps", Itu.int, 2021. [Online]. Available: https://www.itu.int/en/ITU-D/Technology/Pages/InteractiveTransmissionMaps.aspx. [Accessed: 15- Feb- 2021].

\bibitem{b16} E. Oughton, "Policy choices can help keep 4G and 5G universal broadband affordable",\emph{arXiv:2101.07820 [cs, econ, q-fin]}, Jan. 2021, Accessed: Jan. 21, 2021. [Online]. Available: http://arxiv.org/abs/2101.07820. 2021

\bibitem{b17} "Elon, Masa and Boris in low-Earth orbit". [Online]. Available: https://www.economist.com/business/2020/07/09/elon-masa-and-boris-in-low-earth-orbit. [Accessed: 15- Feb- 2021].

\bibitem{b18}Z. Zhang et al., "User Activity Detection and Channel Estimation for Grant-Free Random Access in LEO Satellite-Enabled Internet of Things," \emph{in IEEE Internet of Things Journal}, vol. 7, no. 9, pp. 8811-8825, Sept. 2020, doi: 10.1109/JIOT.2020.2997336.

\bibitem{b19}T. Narytnik, B. Rassamakin, V. Prisyazhny and S. Kapshtyk, "Coverage Aarea Formation for a Low-Orbit Broadband Access System with Distributed Satellites," \emph{2018 International Conference on Information and Telecommunication Technologies and Radio Electronics (UkrMiCo)}, Odessa, Ukraine, 2018, pp. 1-4, doi: 10.1109/UkrMiCo43733.2018.9047526. 

\bibitem{b20}J. A. Ruiz de Azúa, A. Calveras and A. Camps, "Internet of Satellites (IoSat): Analysis of Network Models and Routing Protocol Requirements," \emph{in IEEE Access}, vol. 6, pp. 20390-20411, 2018, doi: 10.1109/ACCESS.2018.2823983. 

\bibitem{b21}Y. Yan, K. An, B. Zhang, W. -P. Zhu, G. Ding and D. Guo, "Outage Constrained Robust Multigroup Multicast Beamforming for Satellite-Based Internet of Things Coexisting With Terrestrial Networks," \emph{in IEEE Internet of Things Journal}, doi: 10.1109/JIOT.2020.3042831.

\bibitem{b22} Oughton, E.J., Lehr, W., Katsaros, K., Selinis, I., Bubley, D. and Kusuma, J., 2021. Revisiting wireless internet connectivity: 5G vs Wi-Fi 6. \emph{Telecommunications Policy}, 45(5), p.102127.

\bibitem{b23} Oughton, E.J. and Russell, T., 2020. The importance of spatio-temporal infrastructure assessment: Evidence for 5G from the Oxford–Cambridge Arc. \emph{Computers, Environment and Urban Systems}, 83, p.101515.

\bibitem{b24} Rendón Schneir, J., Whalley, J., Amaral, T.P., Pogorel, G., 2018. The implications of 5G networks: Paving the way for mobile innovation? Telecommunications Policy, The implications of 5G networks: Paving the way for mobile innovation? 42, 583–586.

\bibitem{b25} E. J. Oughton, K. Konstantinos, E. Fariborz, K. Dritan and C. Jon, “An open-source techno-economic assessment framework for 5G deployment.,” \emph{IEEE Access}, vol. 7, pp. 155930-155940, 2019. 

\bibitem{b26} E. J. Oughton and F. Zoraida, “The cost, coverage and rollout implications of 5G infrastructure in Britain,” \emph{Telecommunications Policy}, vol. 8, no. 42, pp. 636-652, 2018. 

\bibitem{b27} E. J. Oughton, F. Zoraida, v. d. G. Sietse and v. d. B. Rudolf, “Assessing the capacity, coverage and cost of 5G infrastructure strategies: Analysis of the Netherlands.,” \emph{Telematics and Informatics} , vol. 37, pp. 50-69, 2019. 

\bibitem{b28} E. J. Oughton, J. Mathur, Policy options for digital infrastructure strategies: A simulation model for broadband universal service in Africa’, \emph{arXiv:2102.03561 [cs, econ, q-fin]},Feb.2021, Accessed: Feb. 09, 2021. [Online]. Available \underline{https://arxiv.org/abs/2102.03561}

\bibitem{b29} L. Zhen, A. K. Bashir, K. Yu, Y. D. Al-Otaibi, C. H. Foh and P. Xiao, "Energy-Efficient Random Access for LEO Satellite-Assisted 6G Internet of Remote Things," \emph{in IEEE Internet of Things Journal}, doi: 10.1109/JIOT.2020.3030856.

\bibitem{b30} K. S. Rao and D. R. Jahagirdar, "Quadrifilar helical antenna with integrated compact feed for TTC application in LEO satellites," \emph{2018 IEEE Indian Conference on Antennas and Propogation (InCAP)}, Hyderabad, India, 2018, pp. 1-4, doi: 10.1109/INCAP.2018.8770934.

\bibitem{b31} Z. Qu, Z. Gengxin, C. Haotong and X. Jidong, “LEO Satellite Constellation for Internet of Things,” \emph{IEEE Access 5}, pp. 18391-18401, 2017. 

\bibitem{b32} Y. Su, L. Yaoqi, Z. Yiqing, Y. Jinhong, C. Huan and S. Jinglin, “Broadband LEO satellite communications: Architectures and key technologies,” \emph{IEEE Wireless Communications}, vol. 26, no. 2, pp. 55-61, 2019. 

\bibitem{b33} A. Chiha, V. d. W. Marlies, C. Didier and V. Sofie, “Techno-economic viability of integrating satellite communication in 4G networks to bridge the broadband digital divide,” \emph{Telecommunications Policy}, vol. 44, no. 3, pp. 1-17, 2020. 

\bibitem{b34} N. Wang, "Content Delivery for High-Speed Railway via Integrated Terrestrial-Satellite Networks", \emph{2020 IEEE Wireless Communications and Networking Conference (WCNC). IEEE}, 2020. [Accessed 15 February 2021].

\bibitem{b35} R. L. Sturdivant and J. Lee, "Systems engineering of IoT connected commercial airliners using satellite backhaul links," \emph{2018 IEEE Topical Workshop on Internet of Space (TWIOS)}, Anaheim, CA, USA, 2018, pp. 1-4, doi: 10.1109/TWIOS.2018.8311397. 

\bibitem{b36} J. A. Ruiz-De-Azua, V. Ramírez, H. Park, A. C. AUGé and A. Camps, "Assessment of Satellite Contacts Using Predictive Algorithms for Autonomous Satellite Networks," \emph{n IEEE Access}, vol. 8, pp. 100732-100748, 2020, doi: 10.1109/ACCESS.2020.2998049.

\bibitem{b37} H. Tan, M. He, T. Xia, X. Zheng and J. Lai, "A Novel Multi-level Computation Offloading Scheme at LEO Constellation Broadband Network Edge," \emph{2020 IEEE World Congress on Services (SERVICES)}, Beijing, China, 2020, pp. 281-286, doi: 10.1109/SERVICES48979.2020.00062.

\bibitem{b38} L. Wang, S. Liu, W. Wang and Z. Fan, "Dynamic uplink transmission scheduling for satellite Internet of Things applications," \emph{in China Communications}, vol. 17, no. 10, pp. 241-248, Oct. 2020, doi: 10.23919/JCC.2020.10.018.

\bibitem{b39} L. Zhen, A. Bashir, K. Yu, Y. Al-Otaibi, C. Foh and P. Xiao, "Energy-Efficient Random Access for LEO Satellite-Assisted 6G Internet of Remote Things", \emph{IEEE Internet of Things Journal}, pp. 1-1, 2020. Available: 10.1109/jiot.2020.3030856.

\bibitem{b40} J. Dvořák, J. Hart and K. J, “Atmospheric attenuation of the Ku band along the space-earth path due to clouds and rain,” \emph{Agronomy Research}, vol. 18, no. S2, p. 1235–1243, 2020. 

\bibitem{b41} O. Kodheli, N. Maturo, S. Andrenacci, S. Chatzinotas and F. Zimmer, “Link Budget Analysis for Satellite-Based Narrowband IoT Systems,” \emph{In International Conference on Ad-Hoc Networks and Wireless}, pp. 259-271, 2019. 

\bibitem{b42} A. Gaber, M. A. ElBahaay, A. Maher Mohamed, M. M. Zaki, A. Samir Abdo and N. AbdelBaki, "5G and Satellite Network Convergence: Survey for Opportunities, Challenges and Enabler Technologies," \emph{2020 2nd Novel Intelligent and Leading Emerging Sciences Conference (NILES)}, Giza, Egypt, 2020, pp. 366-373, doi: 10.1109/NILES50944.2020.9257914. 

\bibitem{b43} P. A. Iannucci and T. E. Humphreys, "Economical Fused LEO GNSS," \emph{2020 IEEE/ION Position, Location and Navigation Symposium (PLANS)}, Portland, OR, USA, 2020, pp. 426-443, doi: 10.1109/PLANS46316.2020.9110140.

\bibitem{b44} F. Ma et al., "Hybrid constellation design using a genetic algorithm for a LEO-based navigation augmentation system", \emph{GPS Solutions}, vol. 24, no. 2, 2020. Available: 10.1007/s10291-020-00977-0.

\bibitem{b45}Y. Zhao, J. Cao and Y. Li, "An Improved Timing Synchronization Method for Eliminating Large Doppler Shift in LEO Satellite System," \emph{2018 IEEE 18th International Conference on Communication Technology (ICCT)}, Chongqing, China, 2018, pp. 762-766, doi: 10.1109/ICCT.2018.8600170. 

\bibitem{b46} E. Meng and X. Bu, "Two-Dimensional Joint Acquisition of Doppler Factor and Delay for MC-DS-CDMA in LEO Satellite System," \emph{in IEEE Access}, vol. 8, pp. 148203-148213, 2020, doi: 10.1109/ACCESS.2020.3015754. 

\bibitem{b47} S. Liu, X. Hu and W. Wang, "Deep Reinforcement Learning Based Dynamic Channel Allocation Algorithm in Multibeam Satellite Systems", \emph{IEEE Access}, vol. 6, pp. 15733-15742, 2018. Available: 10.1109/access.2018.2809581.

\bibitem{b48} S. Shrestha and D. Choi, "Characterization of Rain Specific Attenuation and Frequency Scaling Method for Satellite Communication in South Korea", \emph{International Journal of Antennas and Propagation}, vol. 2017, pp. 1-16, 2017. Available: 10.1155/2017/8694748.

\bibitem{b49} W. Damascena Dias, M. Carleti, S. Souza Lima Moreira and L. Leonel Mendes, "Evaluation of Rain Attenuation Models in Satellite Links under Tropical and Equatorial Climates," \emph{in IEEE Latin America Transactions}, vol. 16, no. 2, pp. 358-367, Feb. 2018, doi: 10.1109/TLA.2018.8327387. 

\bibitem{b50} M. Marchese, F. Patrone and M. Cello, "DTN-Based Nanosatellite Architecture and Hot Spot Selection Algorithm for Remote Areas Connection," \emph{in IEEE Transactions on Vehicular Technology}, vol. 67, no. 1, pp. 689-702, Jan. 2018, doi: 10.1109/TVT.2017.2739298. 

\bibitem{b51} S. Agnelli, P. Feltz, P. Griffiths and D. Roth, "Satellite's role in the penetration of broadband connectivity within the European Union," \emph{2014 7th Advanced Satellite Multimedia Systems Conference and the 13th Signal Processing for Space Communications Workshop (ASMS/SPSC)}, Livorno, Italy, 2014, pp. 306-311, doi: 10.1109/ASMS-SPSC.2014.6934560. 

\bibitem{b52} M. Cello, M. Marchese and F. Patrone, "HotSel: A Hot Spot Selection Algorithm for Internet Access in Rural Areas through Nanosatellite Networks," \emph{2015 IEEE Global Communications Conference (GLOBECOM)}, San Diego, CA, USA, 2015, pp. 1-6, doi: 10.1109/GLOCOM.2015.7417202. 

\bibitem{b53} E. Oughton, T. Russell, J. Johnson, C. Yardim and J. Kusuma, "itmlogic: The Irregular Terrain Model by Longley and Rice", \emph{Journal of Open Source Software}, vol. 5, no. 51, p. 2266, 2020. Available: 10.21105/joss.02266.

\bibitem{b54} N. Ioannou, D. Katsianis and D. Varoutas, "Comparative techno-economic evaluation of LTE fixed wireless access, FTTdp G.fast and FTTC VDSL network deployment for providing 30 Mbps broadband services in rural areas", \emph{Telecommunications Policy}, vol. 44, no. 3, p. 101875, 2020. Available: 10.1016/j.telpol.2019.101875.

\bibitem{b55} Oughton, E., Lehr, W., 2021. Next-G Wireless: Learning from 5G Techno-Economics to Inform Next Generation Wireless Technologies (SSRN Scholarly Paper No. ID 3898099). \emph{Social Science Research Network}, Rochester, NY.

\bibitem{b56} V. Krizanovic Cik, D. Zagar and K. Grgic, "A framework for optimal techno-economic assessment of broadband access solutions and digital inclusion of rural population in global information society", \emph{Universal Access in the Information Society}, vol. 17, no. 3, pp. 517-540, 2017. Available: 10.1007/s10209-017-0560-x.

\bibitem{b57} T. Vrind, S.  Rao, L. Pathak, and D. Das, ‘Deep Learning-based LAP Deployment and Aerial Infrastructure Sharing in 6G’, \emph{in 2020 IEEE International Conference on Electronics, Computing and Communication Technologies (CONECCT)}, Jul. 2020, pp. 1–5, doi: 10.1109/CONECCT50063.2020.9198319.

\bibitem{b58} A. Tzanakaki, M. P. Anastasopoulos and D. Simeonidou, "Converged optical, wireless, and data center network infrastructures for 5G services," \emph{in IEEE/OSA Journal of Optical Communications and Networking}, vol. 11, no. 2, pp. A111-A122, Feb. 2019, doi: 10.1364/JOCN.11.00A111. 

\bibitem{b59} K. Kanta et al., "Analog fiber-wireless downlink transmission of IFoF/mmWave over in-field deployed legacy PON infrastructure for 5G fronthauling," \emph{in IEEE/OSA Journal of Optical Communications and Networking}, vol. 12, no. 10, pp. D57-D65, October 2020, doi: 10.1364/JOCN.391803. 

\bibitem{b60} I. Akyildiz, A. Kak and S. Nie, "6G and Beyond: The Future of Wireless Communications Systems", \emph{IEEE Access}, vol. 8, pp. 133995-134030, 2020. Available: 10.1109/access.2020.3010896.

\bibitem{b61} Oughton, E.J., Comini, N., Foster, V., Hall, J.W., 2021. Policy Choices Can Help Keep 4G and 5G Universal Broadband Affordable. \emph{World Bank, Policy Research Working Paper 1}.

\bibitem{b62} Y. Firmansyah, N. Rahayu, Y. Prabowo, I. Putro and F. Kurniawan, "Quality of Service (QoS) Analysis for Real-Time Telemetry By IP Satellite Communication", \emph{2020 International Conference on Radar, Antenna, Microwave, Electronics, and Telecommunications (ICRAMET)}, pp. 18-21, 2020. [Accessed 15 February 2021].

\bibitem{b63} T. Vrind, S. Rao, L. Pathak, and D. Das, ‘Deep Learning-based LAP Deployment and Aerial Infrastructure Sharing in 6G’, \emph{in 2020 IEEE International Conference on Electronics, Computing and Communication Technologies (CONECCT)}, Jul. 2020, pp. 1–5, doi: 10.1109/CONECCT50063.2020.9198319. 

\bibitem{b64} A. Gaber, M. A. ElBahaay, A. Maher Mohamed, M. M. Zaki, A. Samir Abdo and N. AbdelBaki, "5G and Satellite Network Convergence: Survey for Opportunities, Challenges and Enabler Technologies," \emph{2020 2nd Novel Intelligent and Leading Emerging Sciences Conference (NILES)}, Giza, Egypt, 2020, pp. 366-373, doi: 10.1109/NILES50944.2020.9257914. 

\bibitem{b65} Z. Yuan, J. Jin, L. Sun, K. Chin and G. Muntean, "Ultra-Reliable IoT Communications with UAVs: A Swarm Use Case", \emph{IEEE Communications Magazine}, vol. 56, no. 12, pp. 90-96, 2018. Available: 10.1109/mcom.2018.1800161.

\bibitem{b66}S. Aggarwal, N. Kumar, and S. Tanwar, ‘Blockchain Envisioned UAV Communication Using 6G Networks: Open issues, Use Cases, and Future Directions’, \emph{IEEE Internet of Things Journal}, pp. 1–1, 2020, doi: 10.1109/JIOT.2020.3020819. 

\bibitem{b67} I. del Portillo, S. Eiskowitz, E. Crawley and B. Cameron, "Connecting the other half: Exploring options for the 50\% of the population unconnected to the internet", \emph{Telecommunications Policy}, vol. 45, no. 3, p. 102092, 2021. Available: 10.1016/j.telpol.2020.102092.

\bibitem{b68} B. Evans, O. Onireti, T. Spathopoulos and M. A. Imran, "The role of satellites in 5G," \emph{2015 23rd European Signal Processing Conference (EUSIPCO)}, Nice, France, 2015, pp. 2756-2760, doi: 10.1109/EUSIPCO.2015.7362886.

\bibitem{b69} C. Ge et al., "QoE-Assured Live Streaming via Satellite Backhaul in 5G Networks," \emph{in IEEE Transactions on Broadcasting}, vol. 65, no. 2, pp. 381-391, June 2019, doi: 10.1109/TBC.2019.2901397. 

\bibitem{b70} J. Kim et al., "5G-ALLSTAR: An Integrated Satellite-Cellular System for 5G and Beyond," \emph{2020 IEEE Wireless Communications and Networking Conference Workshops (WCNCW)}, Seoul, Korea (South), 2020, pp. 1-6, doi: 10.1109/WCNCW48565.2020.9124751. 

\bibitem{b71} K. Liolis, G. Alexander, S. Ray, S. Detlef, W. Simon, P. Georgia, E. Barry, W. Ning, V. Oriol, T. J. Boris, F. Michael, D. S. Salva, S. K. Pouria and C. Nicolas, “Use cases and scenarios of 5G integrated satellite‐terrestrial networks for enhanced mobile broadband: The SaT5G approach.,” \emph{International Journal of Satellite Communications and Networking}, vol. 37, no. 2, pp. 91-112, 2019. 

\bibitem{b72} I. del Portillo, B. Cameron and E. Crawley, "A technical comparison of three low earth orbit satellite constellation systems to provide global broadband", \emph{Acta Astronautica}, vol. 159, pp. 123-135, 2019. Available: 10.1016/j.actaastro.2019.03.040.

\bibitem{b73} J. Galán, L. Chiaraviglio, L. Amorosi and N. Blefari-Melazzi, "Multi-Period Mission Planning of UAVs for 5G Coverage in Rural Areas:", In 2018 9th International Conference on the Network of the Future (NOF), pp. 52-59, 2018. [Accessed 15 February 2021].

\bibitem{b74} Oughton, E.J., Russell, T., 2020. cdcam: Cambridge Digital Communications Assessment Model. \emph{Journal of Open Source Software 5}, 1911.

\bibitem{b75} J. Hall et al., "Strategic analysis of the future of national infrastructure", \emph{Proceedings of the Institution of Civil Engineers-Civil Engineering}, vol. 170, no. 1, pp. 39-47, 2021.

\bibitem{b76} S. F. Yunas, V. Mikko and N. Jarno, “Spectral and energy efficiency of ultra-dense networks under different deployment strategies,” \emph{EEE Communications Magazine}, vol. 53, no. 1, pp. 90-100, 2015.

\bibitem{b77} E. Olakunle, T. Abdul Rahman, H. Saharuddin and F. Khairodin, "Factors that Impact LoRa IoT Communication Technology", 2019 IEEE 14th Malaysia International Conference on Communication (MICC), pp. 112-117, 2019. [Accessed 15 February 2021].

\bibitem{b78} European Telecommunications Standards Institute, Digital Video Broadcasting (DVB); Second generation framing structure, channel coding and modulation systems for Broadcasting, Interactive Services, News Gathering and other broadband satellite applications;. Sophia-Antipolis, 2021.

\bibitem{b79} Oughton, E.J., Jha, A., 2021. Supportive 5G Infrastructure Policies are Essential for Universal 6G: Assessment Using an Open-Source Techno-Economic Simulation Model Utilizing Remote Sensing. \emph{IEEE Access 9}, 101924–101945.

\bibitem{b80} Thoung, C., Beaven, R., Zuo, C., Birkin, M., Tyler, P., Crawford-Brown, D., Oughton, E.J., Kelly, S., 2016. Future demand for infrastructure services, in: The Future of National Infrastructure: A System-of-Systems Approach. \emph{Cambridge University Press, Cambridge}.

\bibitem{b81} "WorldPop :: Population Counts", Worldpop.org, 2021. [Online]. Available: \underline{https://www.worldpop.org/project/categories?id=3}. [Accessed: 03- May- 2021].

\bibitem{b82} J. Hindin, TECHNICAL APPENDIX: Application of Kuiper Systems LLC for Authority to Launch and Operate a Non-Geostationary Satellite Orbit System in Ka-band Frequencies. Washington DC: \emph{Federal Communications Commission}, 2019.

\bibitem{b83} Federal Communications Commission, SpaceX Non-Geostationary Satellite System Attachment a Technical Information to Supplement Schedule S. Washington DC: https://fcc.report/IBFS/SAT-MOD-20181108-00083/1569860.pdf, 2021.

\bibitem{b84} SES, \emph{SES Annual Report 2019.} Luxembourg City,  Luxembourg:, Accessed on: Feb. 16, 2021, [Online] Available:{https://www.ses.com/investors/annual-reports}

\bibitem{b85}Chaudhry, A. and Yanikomeroglu, H., 2021. Laser Intersatellite Links in a Starlink Constellation: A Classification and Analysis. \emph{IEEE Vehicular Technology Magazine}, 16(2), pp.48-56. 

\bibitem{b86}Sheetz, M., 2020. Elon Musk touts low cost to insure SpaceX rockets as edge over competitors. CNBC, [online] Available at: \emph{<https://www.cnbc.com/2020/04/16/elon-musk-spacex-falcon-9-rocket-over-a-million-dollars-less-to-insure.html>} [Accessed 25 July 2021]. 

\bibitem{b87}Jones, H., 2018. The Recent Large Reduction in Space Launch Cost. \emph{48th International Conference on Environmental Systems}, 48, pp.8-12. 

\bibitem{b88} "GADM maps and data", Global Administrative Areas Database, 2021. [Online]. Available: \underline{https://gadm.org/}. [Accessed: 03- May- 2021].

\bibitem{b89} E. Oughton and B. Osoro, "Global Satellite Assessment Tool (globalsat)", GitHub, 2021. [Online]. Available: \underline{https://github.com/edwardoughton/globalsat}. [Accessed: 06- May- 2021].

\end{thebibliography}
\end{document}